\documentclass[a4paper]{article}

\usepackage[a4paper,top=3cm,bottom=2cm,left=2.5cm,right=2.5cm,marginparwidth=1.75cm]{geometry}
\usepackage[caption=false,font=footnotesize]{subfig}
\usepackage{times}
\usepackage{amssymb,amsmath}
\usepackage{graphicx}
\usepackage[lined,ruled,commentsnumbered,linesnumbered]{algorithm2e}
\usepackage{color}
\usepackage{url}
\usepackage{breakurl}

\usepackage{placeins}
\usepackage{etoolbox}
\usepackage{comment}
\usepackage{blindtext}
\usepackage{multirow}
\usepackage{scrextend}
\addtokomafont{labelinglabel}{\sffamily}
\usepackage[official]{eurosym}
\usepackage{caption}
\usepackage{float}
\usepackage{color}
\usepackage{cite}
\usepackage{amssymb,amsmath}
\usepackage{graphicx}
\usepackage[lined,ruled,commentsnumbered,linesnumbered]{algorithm2e}
\usepackage{color}
\usepackage{url}
\usepackage{placeins}
\usepackage{etoolbox}
\usepackage{comment}
\usepackage{blindtext}
\usepackage{multirow}
\usepackage{scrextend}
\addtokomafont{labelinglabel}{\sffamily}
\usepackage[official]{eurosym}
\usepackage{cite}

\begin{document}
\pagenumbering{gobble}


\title{Frictionless Authentication System: Security \& Privacy Analysis and Potential Solutions}

\author{Mustafa~A.~Mustafa,~Aysajan~Abidin,~and~Enrique~Argones~R\'{u}a
  \thanks{This work was supported by the Research Council KU Leuven: C16/15/058, the European Commission through the SECURITY programme under FP7-SEC-2013-1-607049 EKSISTENZ, imec through ICON DiskMan, and FWO through SBO SPITE S002417N.}           
 \thanks{M.A. Mustafa, A. Abidin and E. Argones R\'{u}a are with the imec-COSIC research group, Departement of Electrical Engineering (ESAT), KU Leuven, Belgium. e-mail: (\{mustafa.mustafa, aysajan.abidin, enrique.argonesrua\}@esat.kuleuven.be).}
}
\date{\vspace{-5ex}}
\maketitle

\maketitle

\begin{abstract}

This paper proposes a frictionless authentication system, provides a comprehensive security analysis of and proposes potential solutions for this system. It first presents a system that allows users to authenticate to services in a frictionless manner, i.e., without the need to perform any particular authentication-related actions. Based on this system model, the paper analyses security problems and potential privacy threats imposed on users, leading to the specification of a set of security and privacy requirements. These requirements can be used as a guidance on designing secure and privacy-friendly frictionless authentication systems. The paper also sketches three potential solutions for such systems and highlights their advantages and disadvantages.
\end{abstract}


\section{Introduction}
\label{Introduction}

The widespread adoption of mobile and wearable devices by users results in more personal information being stored or accessed using Personal Devices (PDs) such as smartphones. In addition to enhancing user experience, this also creates new opportunities for both users and Service Providers (SPs). However, this also brings with it new security and privacy challenges for both users and SPs~\cite{Sagiroglu2013}. Usually, these PDs and wearable devices, from now on named Dumb Devices (DDs), have limited computational and interaction capabilities. Nevertheless, users expect a frictionless user experience (making minimum effort) when using their PDs or DDs to access services or resources.  Since these devices are small, light, and easy to carry, they are susceptible to loss and theft, and easier to break. And the use of context information (such as the user's current location, his typical behaviour, etc.), which can easily be accessed from these devices, also triggers privacy concerns. Taking into account the users' needs and the associated security and privacy risks of using such devices, the way users are authenticated and granted access to a wide range of on-line services and content becomes more challenging~\cite{hamme2017}. 

The current authentication systems~\cite{Bhargav-Spantzel2006,Bonneau2012,Grosse2013,Guidorizzi2013,Preuveneers2015} do not provide a satisfactory answer to address these (conflicting) needs: (i) users prefer a single password-less solution, (ii) wearable devices do not offer convenient authentication interface for passwords, (iii) strong biometric authentication solutions score low on usability, or are not suited for continuous authentication with minimal interaction with the user, (iv) certain risk-based techniques work well for desktop and laptops (e.g., device fingerprints), but fall short on mobile devices, and (v) smartphones and wearables are more prone to loss and theft. Thus, there is a clear need for solutions that are tailored towards the user, his devices, the context and sensitivity of his assets.


In this paper, we propose a Frictionless Authentication System (FAS) that allows users to authenticate themselves using their devices to third party SPs without intentionally performing any authentication-related specific actions. We also analyse the security and privacy implications of such systems and propose three potential solutions. The main contributions of this paper are three-fold.

\begin{itemize}
    
\item[-] Firstly, it proposes a novel FAS that allows secure, privacy-friendly as well as frictionless user experience when a user authenticates to SPs.
    
\item[-] Secondly, it performs a threat analysis of and specifies a set of security and privacy requirements for the FAS.
   
\item[-] Thirdly, it proposes three potential high level solutions to achieve secure and privacy-friendly FAS.

\end{itemize}

The remainder of this paper is organised as follows: Section~\ref{Related Work} discusses related work. Section~\ref{Frictionless Authentication System} proposes a frictionless authentication system. Section~\ref{Threat Analysis} analyses potential security threats and attacks to the proposed system. Section~\ref{Security and Privacy Requirements} specifies a set of security and privacy requirements. Section~\ref{Potential Solutions} provides a high-level overview of three potential solutions for a secure and privacy-friendly FAS. Finally, Section~\ref{Conclusions} concludes the paper.

\section{Related Work}
\label{Related Work}

In contrast to conventional challenge-response protocols which use a single prover and verifier, collaborative authentication schemes use a challenge-response protocol with multiple collaborating provers and a single verifier. To mitigate the threat of PDs/DDs being stolen or lost as well as to support a dynamic set of devices as users may not always carry the same set, threshold-based cryptography is used. Threshold cryptography allows one to protect a key by sharing it amongst a number of devices in such a way that (i) only a subset of the shares with minimal size (a threshold $t+1$) can use the key and (ii) having access to $t$ or less shares does not leak any information about the key. Shamir~\cite{Sha79} first introduced this concept of secret sharing, which was later extended to verifiable secret sharing by~Feldman~\cite{Feldman1987}. Pedersen~\cite{Pedersen92} used this concept to construct the first Distributed Key Generation (DKG) protocol. Shoup~\cite{shoup2000} showed how to transform a standard signature scheme such as RSA into a threshold-based variant. In 2010, Simoens~et~al.~\cite{Simoens2010} presented a new DKG protocol which allows devices not capable of securely storing secret shares to be incorporated into threshold signature schemes. Peeters~et~al.~\cite{Peeters2012} proposed a threshold-based distance bounding protocol which also takes into account the proximity of devices holding the share to the verifier. An overview of recent developments in continuous authentication schemes is given in~\cite{patel2016continuous}.


\section{Frictionless Authentication System}
\label{Frictionless Authentication System}

This section details the system model, functional requirements, and interactions amongst entities of a FAS.


\begin{figure}[t]
\centering
\includegraphics[width=0.8\textwidth]{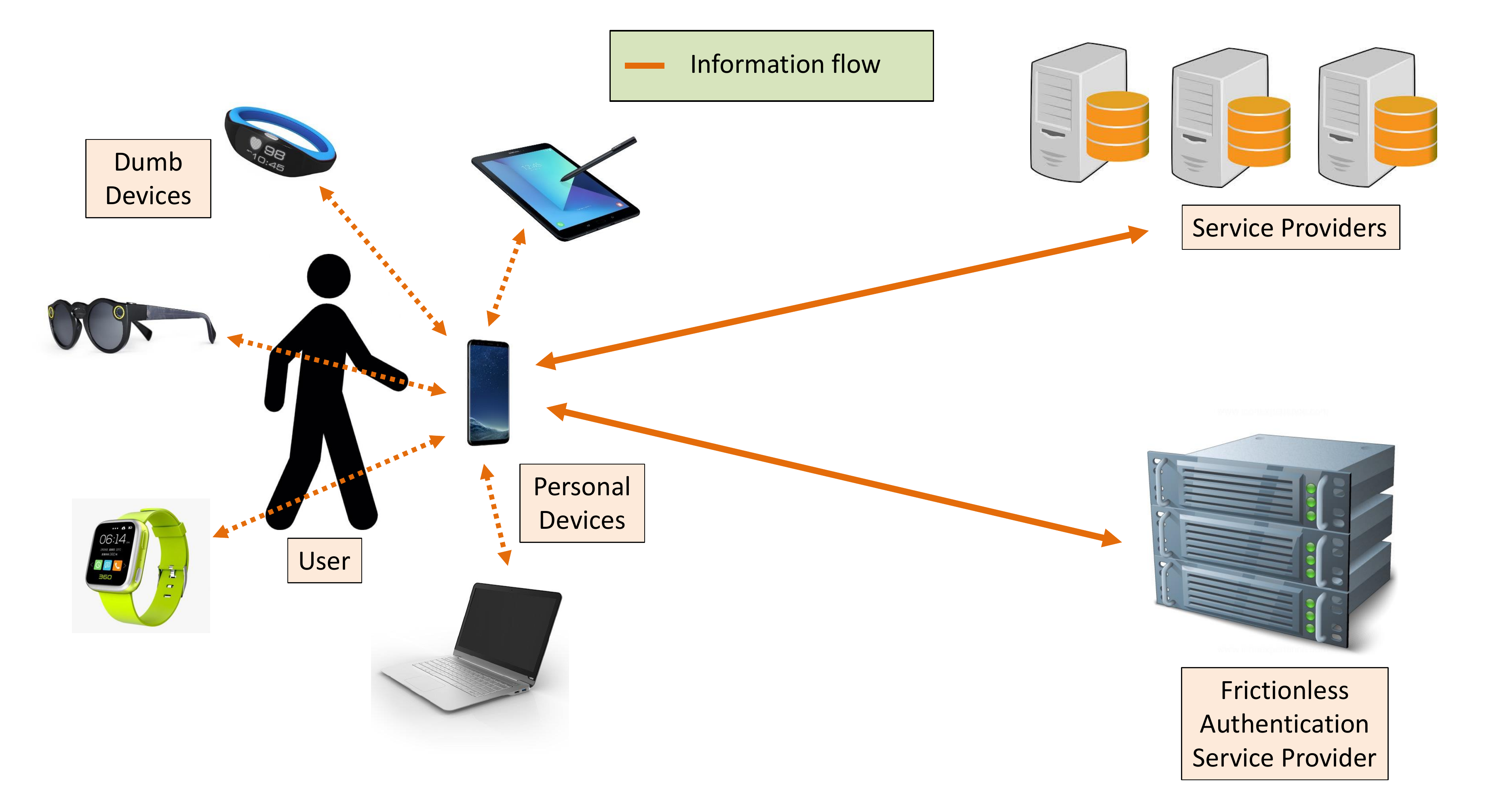}
\caption{A system model of a FAS.}
\label{Fig:SystemModel}
\end{figure}


\subsection{A System Model}

As shown in Figure~\ref{Fig:SystemModel}, a system model of a FAS consists of the following entities. A \textit{user} who wants to access various services provided by different \textit{Service Providers (SPs)}. The user also carries or wears a number of personal or dumb devices which she uses to authenticate herself in a frictionless manner, i.e., without intentionally performing any specific authentication-related actions such as entering a password. \textit{Personal Devices (PDs)} are owned by the user and have a secure storage where their owner's secret data such as (parts of) her private key can be stored. The user communicates with SPs via her PDs. \textit{Dump Devices (DDs)} do not have secure storage. They can communicate with PDs, but not necessarily with the SPs. Usually, DDs are wearable which are not paired with the user, and have sensors. Each PD and DD may have one or more \textit{sensors} integrated to measure different data such as location, gait, blood pressure and heart beats. \textit{SPs} are the entities to which users want to authenticate in order to have access to data or services. Usually, this authentication is done by a user digitally signing a challenge sent by the SP. \textit{Frictionless Authentication Service Provider (FASP)} is the SP that assists users in performing a frictionless authentication.


\subsection{Functional Requirements}

To be practical and adopted by users, any FAS should be: \textit{frictionless} - the involvement of the user should be minimum while authenticating to various SPs; \textit{adaptive} - the FASP should be able to tailor the multi-modal and -factor authentication scheme to user content data; \textit{collaborative} - the authentication score (AuthScore), i.e., the score which determines how confident the FASP is that the user is who she is claiming to be, should be constructed based on data provided by multiple user's PDs and/or DDs; \textit{flexible} - AuthScore should be constructable using various combinations of user's data collected by user's PDs/DDs; \textit{robust \& resilient} - a failure/lack of a single user device should not require any additional effort by the user; and \textit{compatible} - a user should always be able to use conventional authentication methods if desired or needed.

\subsection{Interactions among Entities}

Next, we describe the potential message types and interactions among the entities within the FAS. 

\textbf{System setup.} The FASP performs all the necessary initial steps in order to assist users experience frictionless authentication service. These steps include obtaining the necessary cryptographic keys and certificates.

\textbf{User device setup/registration.} The user obtains or generates a public/private key pair and a certificate for the public key. The entire (or part of the) private key is stored in her PDs. 

\textbf{User registration.} A user provides the SPs with all the necessary information for the service registration such as user identity, public key and certificate.

\textbf{Frictionless authentication.} The user proves her identity to a SP without performing any intentional authentication-related actions. It consists of the following five steps.

\begin{itemize}

\item \textit{Authentication request:} a user informs a SP that she wants to access data or service provided by the SP, or the SP informs the user that she will have to prove her identity.

\item \textit{Identity verification challenge:} the SP sends a challenge to the user to prove her identity. 

\item \textit{User AuthScore calculation:} a user's data gathered by the user's PDs and/or DDs are forwarded (via a single user PD) to the FASP which uses these data to compute the AuthScore of the user. Such calculation could be performed on demand (when requested by the SP) or continuously. If the AuthScore is above a certain predefined threshold, the user's private key becomes available for use. Note that the AuthScore can be computed by the FASP on the cloud or locally on the user's PD. See Section~\ref{cloud v local} for more details regarding the choice of where the AuthScore is calculated. 

\item \textit{Identity verification response:} the user uses her private key to digitally sign the verification challenge and sends the result to the SP. 

\item \textit{User identity verification and service access:} the SP checks the user response and if the verification response holds, it grants the user with access to the requested data or services. 

\end{itemize}


\section{Threat Analysis}
\label{Threat Analysis}

We describe the threat model and provide an analysis of the security and privacy threats to the proposed FAS.




\subsection{Threat Model}

\textit{Users are untrustworthy and malicious.} A malicious user might try passively and/or actively to collect and alter the information stored and exchanged within the FAS, in an attempt to gain access to data or services which she does not have permission to access. \textit{PDs are trustworthy (tamper-evident).} We assume that PDs are equipped with security mechanisms to provide access control and protection against data breaches and/or malware. \textit{DDs are untrustworthy.} The data they measure and forward to the FASP might be corrupted. \textit{The FASP is honest-but-curious.} It follows the protocol specification, but it might try to learn and extract unauthorised information about users. \textit{SPs are untrustworthy or even malicious.} They may try to eavesdrop and collect information exchanged within the FAS. Their aim might be to gain access, collect and/or modify information exchanged within a FAS in an attempt to disrupt, and extract confidential information about users, competitors (other SPs) and the FASP itself. 






\subsection{Security Analysis}

This section analyses the possible security threats to a FAS. The analysis is based on the STRIDE framework~\cite{microsoft_web_app_sec} which mainly covers security threats.

\begin{itemize}

\item \textit{Spoofing:} A malicious entity may attempt to get unauthorised access to services provided by a SP. Such spoofing attacks introduce trust related issues, and may have an economic impact to the SP, especially if the SP provides financial services. Hence, it is important to have thorough user registration procedures and strong mutual authentication.

\item \textit{Tampering with data:} A malicious entity may attempt to modify the information stored and/or exchanged within the FAS such as manipulating (i) the data sent from the sensors of a user's devices, (ii) AuthScore and/or (3) user content data such as location. By stating inaccurate information, an adversary may attempt to lower the credibility of users, SPs and the FASP. Therefore, the integrity and authenticity of the data exchanged/stored should be guaranteed.

\item \textit{Information disclosure:} A malicious entity may attempt to eavesdrop messages sent within the FAS. By eavesdropping messages one may attempt to retrieve information such as who, when, how often and which services access. Such information is considered as private. Hence, confidentiality of data must be guaranteed. Information disclosure also constitutes a privacy threat to users posing additional risks such as users' profiling.

\item \textit{Repudiation:} Disputes may arise when users (do not) access services offered by the SP and claim the opposite. Hence, the non-repudiation of messages exchanged and actions performed by the FAS's entities must be guaranteed, using mechanisms to ensure that disputes are promptly resolved.

\item \textit{Denial-of-Service (DoS):} DoS attacks aim to make the FAS inaccessible to specific or all users. An adversary may target a user's PDs/DDs or the FASP in an attempt to make the service unavailable to that specific user or all users, respectively.

\item \textit{Elevation of privilege:} An adversary may attempt to gain elevated access to SP resources. For instance, a malicious user may attempt to elevate her privileges from accessing the basic available service to accessing premium service, by, for example, manipulating her AuthScore. Thus, to mitigate these attacks, authorization mechanisms that comply with the principle of least privilege should be deployed.

\end{itemize}


\subsection{Privacy Analysis}

This section analyses the possible privacy threats to a FAS. The analysis is based on the LINDDUN framework~\cite{DBLP:journals/re/DengWSPJ11} which mainly covers privacy threats.

\begin{itemize}

\item \textit{Linkability:} An adversary may attempt to distinguish whether two or more Items of Interest (IOI) such as messages, actions and subjects are related to the same user. For instance, an adversary may try to correlate and deduce whether a user has accessed a particular service by a SP at a particular location. Hence, unlikability among IOIs should be guaranteed.

\item \textit{Identifiability:} An adversary may attempt to correlate and identify a user from the types of messages exchanged and actions performed within the FAS. For instance, an adversary may try to identify a user by analysing the messages the user exchanges with the SPs. If a user has considerably more PDs/DDs, this may make her more identifiable. Thus, the anonymity and pseudonymity of users should be preserved.

\item \textit{Non-repudiation:} In contrast to security, non-repudiation can be used against users' privacy. An adversary may attempt to collect evidence stored and exchanged within the FAS to deduce information about a user. It may deduce whether a user has accessed a particular service at a particular location. Thus, plausible deniability over non-repudiation should be provided.

\item \textit{Detectability:} An adversary may try to distinguish the type of IOIs such as messages exchanged amongst FAS entities from a random noise. For instance, an adversary may attempt to identify when a user's PD communicates with a SP. Thus, user undetectability and unobservability should be guaranteed.

\item \textit{Information disclosure:} An adversary may eavesdrop and passively collect information exchanged within the FAS aiming at profiling users. For instance, an adversary may attempt to learn the location and availability of a user. Moreover, the user's behaviour may be inferred by a systematic collection of the user's information~\cite{uber_tracing}. For instance, if a SP and/or the FASP collect the data from the user's PDs/DDs and analyse these data, they may infer (i) the user's health related data by collecting their physiological information, (ii) users' activities by analysing the history of service access, and (iii) circles of trust by analysing with whom, when and how often they use the service. Profiling constitutes a high risk for users' privacy. Thus, the confidentiality of information should be guaranteed.

\item \textit{Content Unawareness:} A misbehaving FASP may attempt to collect more user information than it is necessary aiming to use such information for unauthorised purposes such as advertisement. For instance, the FASP may only need to know whether a user is eligible to access a service without necessarily the need to identify the user nor the service. Hence, the content awareness of users should be guaranteed.

\item \textit{Policy and Consent Noncompliance:} A misbehaving FASP may attempt to collect, store and process users' personal information in contrast to the principles (e.g., data minimisation) described in the European General Data Protection Regulation 2016/680~\cite{regulation2016/680}. For instance, a misbehaving FASP may attempt to (i) collect sensitive information about users such their location, (ii) export users' information to data brokers for revenue without users' consent, and (iii) read users' contacts from their PDs. Thus, privacy policies and consent compliance should be guaranteed.

\end{itemize}

\subsection{Local versus Cloud-based Frictionless Authentication}
\label{cloud v local}

The AuthScore, as mentioned earlier, can be computed by the FASP either on the cloud or locally on a PD of a user. The choice will inevitably affect not only the performance of a FAS but also the risk of privacy breaches. 

\subsubsection{Cloud-based AuthScore Calculation}

The cloud-based AuthScore calculation requires that all user data gathered by the sensors of the user's PDs/DDs are sent to the cloud where the FASP fuses them to compute the AuthScore of the user. Although outsourcing all the calculations to the cloud should allow the FASP to use more complex fusing algorithms, it also adds an additional risk to users' privacy. As some of these data will be highly user-specific, the confidentiality of these data should be protected. In other words, the communication channels between the user's PD and the FASP servers should be encrypted so that no external entity has access to these data. Also, user's privacy should also be protected from the FASP. Having access to these data may allow the FASP to extract sensitive information about the user, thus profiling users. Ideally, the FASP should not have access to the user data in the cleartext, but operate only with encrypted data. This could be achieved if the user data are encrypted with a cryptographic scheme that supports homomorphic properties such as the Paillier cryptosystem~\cite{Paillier}. Moreover, the FASP should not be able to identify the SP to whom the user authenticates. Otherwise, the FASP would be able to track the user online over the different data/services the user accesses.  

\subsubsection{Locally AuthScore Calculation} 

In contrast to the cloud-based solution, calculating the AuthScore on the user's PD is more privacy-friendly as no user data leave the PD. However, on one hand, given that the computational resources of PDs are usually much lower than the ones of the cloud, the complexity of the fusion algorithm will be limited. On the other hand, as the user data is not sent to the FASP services, the fusing algorithm running on the user's PD could use much more fine-grained user data. Having access to such data should allow the FASP to use less complex fusion algorithms but yet achieve results comparable to the ones achieved with more complex fusion algorithms used in cloud-based AuthScore calculation.  


\section{Security and Privacy Requirements}
\label{Security and Privacy Requirements}

Based on the threat analysis, this section specifies a set of security and privacy requirements for the proposed FAS.

\subsection{Security Requirements}\label{sec_req}

To mitigate the aforementioned security threats, the following security requirements needs to be satisfied.

\begin{itemize}

\item \textit{Entity Authentication} assures to an entity that the identity of a second entity is the one that is claiming to be. It aims to mitigate spoofing attacks.

\item \textit{Integrity} ensures that the information stored and exchanged within the FAS have not been altered. It aims to mitigate tampering with data attacks. Integrity is achieved with the use of hash functions, MACs and digital signatures.

\item \textit{Confidentiality} ensures that only the intended entities are able to read the user data stored and transferred within the FAS. It aims to mitigate information disclosure attacks. Confidentiality can be achieved with the use of encryption schemes, e.g., symmetric, asymmetric and homomorphic encryption schemes. 

\item \textit{Non-repudiation} is achieved when an entity cannot deny her action or transaction. It aims to mitigate repudiation attacks (disputes). Non-repudiation can be achieved with the use of digital signatures, timestamps and audit trails.

\item \textit{Availability} ensures that the resources of the FAS are available to legitimate users. It aims to mitigate DoS attacks. To safeguard availability, network tools such as firewalls, intrusion detection and prevention systems should be used. 

\item \textit{Authorisation} ensures that an entity has the correct access. It aims to mitigate elevation of privilege attacks. For authorisation, access control mechanisms, e.g., access control lists and role based access control, should be used, following the principle of least privilege for user accounts.

\end{itemize}

\subsection{Privacy Requirements}\label{pri_req}
To mitigate the specified privacy threats, the following privacy requirements need to be satisfied.

\begin{itemize}

\item \textit{Unlinkability} ensures that two or more IOIs such as messages and actions are not linked to the same user~\cite{pfitzmann2010terminology}. It aims to mitigate linkability attacks. Unlinkability can be achieved with the use of pseudonyms as in~\cite{Mustafa2014}, anonymous credentials~\cite{camenisch2004signature} and private information retrieval~\cite{chor1998private}.

\item \textit{Anonymity} ensures that messages exchanged and actions performed can not be correlated to a user's identity. It aims to mitigate identifiability attacks. Anonymity can be achieved using Mix-nets~\cite{chaum1981untraceable} and multi-party computation. 

\item \textit{Pseudonymity} ensures that a pseudonym is used instead of a user's real identity. As anonymity, it aims to mitigate identifiability attacks. It can be achieved by using unique and highly random data strings as pseudonyms.

\item \textit{Plausible deniability} over non-repudiation ensures that an adversary cannot prove that a user has performed a specific action and operation. It aims to mitigate non-repudiation privacy threats. However, non-repudiation service should be provided when necessary such as when a user needs to be hold accountable for cheating and/or misbehaving, as in~\cite{Symeonidis2017}.

\item \textit{Undetectability and unobservability} ensures that messages exchanged and actions performed by a user cannot be distinguished from others. It aims to mitigate detectability attacks, and can be achieved by using Mix-nets and dummy traffic~\cite{chaum1981untraceable}.

\item \textit{Confidentiality} is a privacy requirement too (see Sect.~\ref{sec_req}).

\item \textit{Content Awareness} aims to raise users' awareness by better informing them of the amount and nature of data they provide the FASP. It aims to mitigate the content unawareness threats, and can be achieved with the use of transparency enhancing technologies, e.g., privacy nudges~\cite{wang2013privacy} and dashboards~\cite{nebel2013personal}. 

\item \textit{Policy and consent compliance} ensures the compliance of the FAS with legislations, e.g., the European General Data Protection Regulation 2016/680~\cite{regulation2016/680}. It aims to mitigate the policy and consent non-compliance privacy threats, and can be achieved with the use of Data Protection Impact Assessments~\cite{dpia} and Privacy Impact Assessments~\cite{wright2011privacy} for the FAS. 

\end{itemize}


\section{Potential Solutions}
\label{Potential Solutions}
In this section, we propose three possible solutions for a FAS and analyse their pros and cons with respect to their security and privacy properties. The authentication is achieved using a digital signature, wherein the private key is held by the user (i.e., the user device) and the verifier (i.e., the SP) challenges the user to prove that she holds the private key by asking her to sign a challenge. However, the solutions differ from each other in the way the private key is handled. 

\subsection{CASE 1: using no Advanced Crypto}

\subsubsection{High-level Description}

\begin{figure}[t]
\centering
\includegraphics[width=0.8\textwidth]{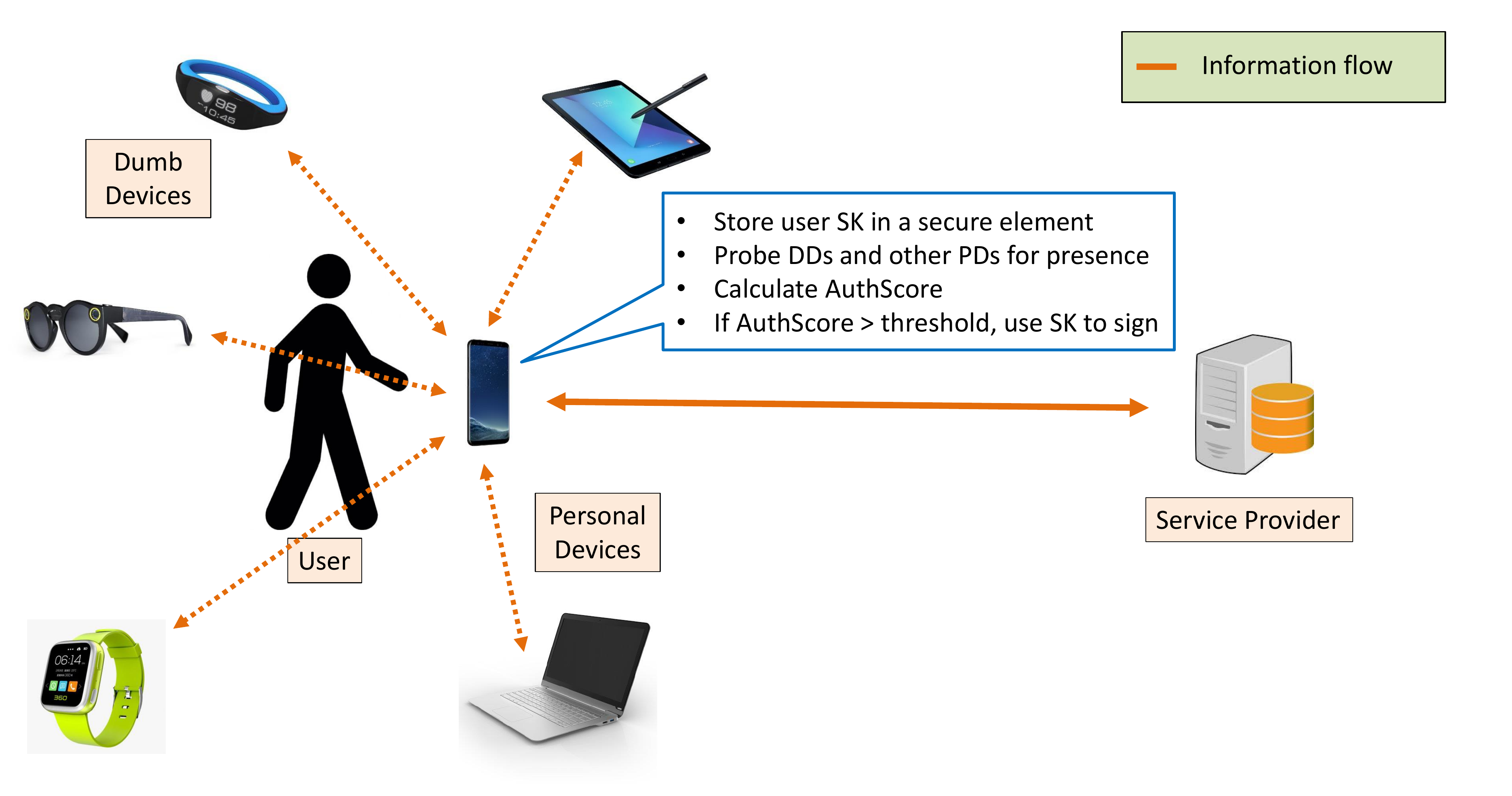}
\caption{CASE 1: FAS using no advanced crypto.}
\label{Fig:case1}
\end{figure}

The first straightforward solution is to password protect the private key. However, this has the obvious drawback of frequent user interaction, as the user has to provide her password every time there is an authentication request. Similarly, protecting the private key using biometrics, e.g., the private key is generated from user biometrics or a local biometric verification is used to grant access to the private key, has the same drawback as the password protected solution. Nevertheless, the user should always be able to authenticate herself using passwords/biometrics. To make it frictionless, one can incorporate behaviometrics/contextual data such as gait, location, or other sensor data. In this case, access to the private key is granted if the behaviometric/contextual data collected from PDs and DDs provide sufficient authentication score; see Figure~\ref{Fig:case1} for a high level description. As can be seen, this solution does not use any advanced cryptographic techniques. 


\subsubsection{Advantages} 
As this solution does not require implementation of any advanced cryptographic algorithms other than the already implemented digital signature algorithm, it is \emph{easy to set-up and implement}. It also has a \emph{simple access control mechanism} as it only requires device presence check and calculation of the AuthScore by matching sensor data. 

\subsubsection{Disadvantages} 
As the key is stored on a single device, this results in \emph{a single point of failure}. Moreover, there are potentially \emph{higher risks for privacy breach} depending on where the AuthScore is calculated based on the behaviometric/contextual data and whether these data are protected.

\subsection{CASE 2: using Threshold Signature}

\subsubsection{High-level Description} 
\begin{figure}[t]
\centering
\includegraphics[width=0.8\textwidth]{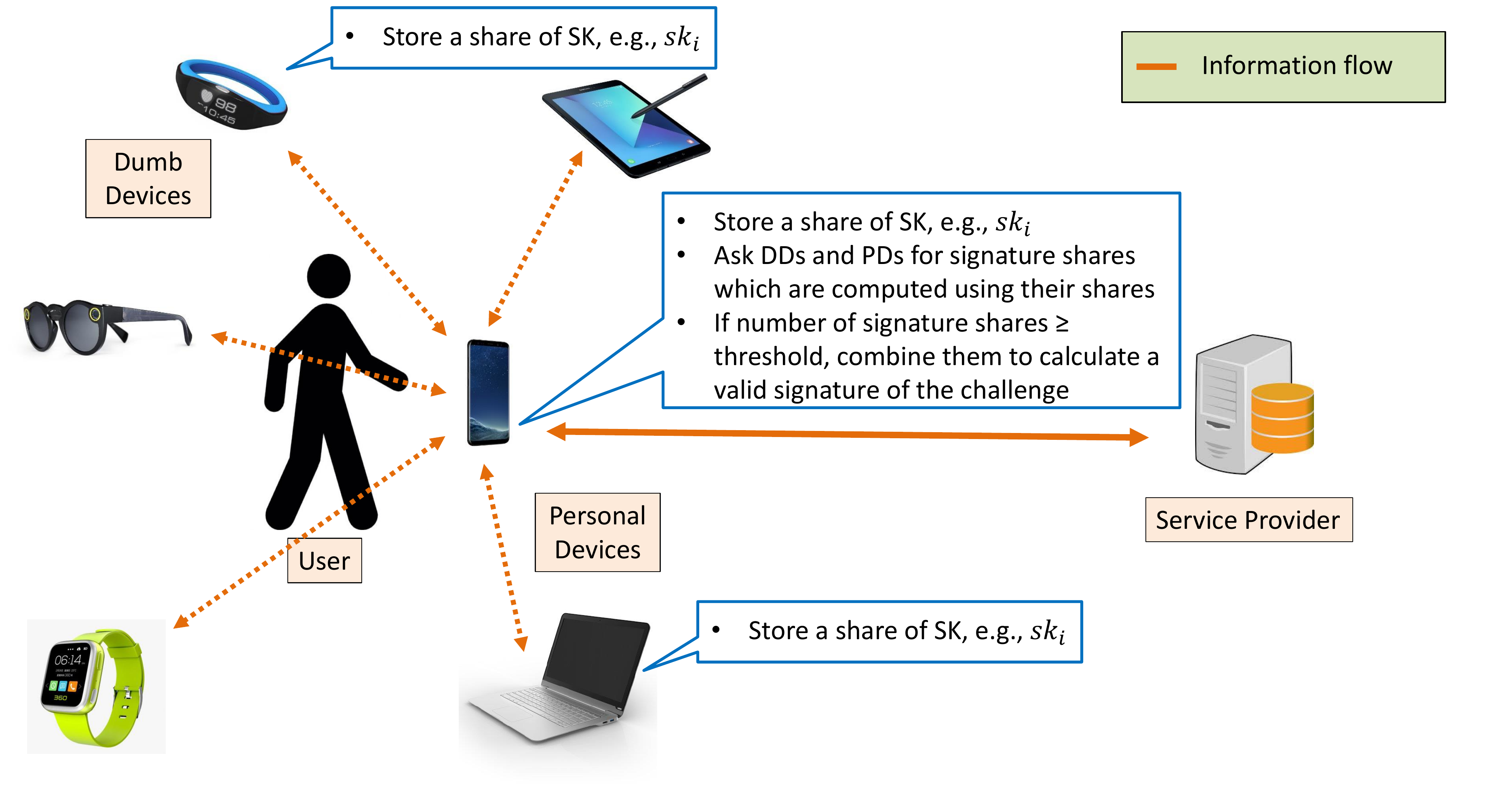}
\caption{CASE 2: FAS using threshold signature.}
\label{Fig:case2}
\end{figure}

The disadvantages of the previous solution can be addressed by using threshold cryptosystems, in particular, threshold signatures~\cite{shoup2000}, as depicted in Figure~\ref{Fig:case2}. In this case, during the enrolment stage, the secret key (i.e., the private key) is shared among the user devices using a threshold secret sharing scheme, so each device stores only a share of the secret key. During the authentication stage, the devices jointly computes a signature on the authentication challenge. In particular, each device computes only one signature share and provides this share to the gateway device, e.g., the user's PD. A valid signature can be computed only if the number of signature shares provided is greater than or equal to a predefined threshold value.  

\subsubsection{Advantages}
As the secret key is shared amongst the user devices and never stored as one piece on any user device, \emph{no key is stored as whole}. Furthermore, the key is not even reconstructed. Only if a sufficiently large enough number of shares (more than the predefined threshold) are stolen, then the key can be reconstructed. Also, as the key is not stored in its entirety, this solution has \emph{no single point of failure}. 

\subsubsection{Disadvantages}
As threshold signatures are more involved than the traditional digital signatures, they may incur some \emph{performance issues in practice}. In addition, even though the key is never stored as a whole, it can be reconstructed using sufficient number of shares. Therefore, \emph{shares need to be protected}. This might be an issue especially for DDs as they usually do not have the capacity for secure storage, which brings us to our third solution described next.


\subsection{CASE 3: using Threshold Signature and Fuzzy Extractors}

\subsubsection{High-level Description}

\begin{figure}[t]
\centering
\includegraphics[width=0.80\textwidth]{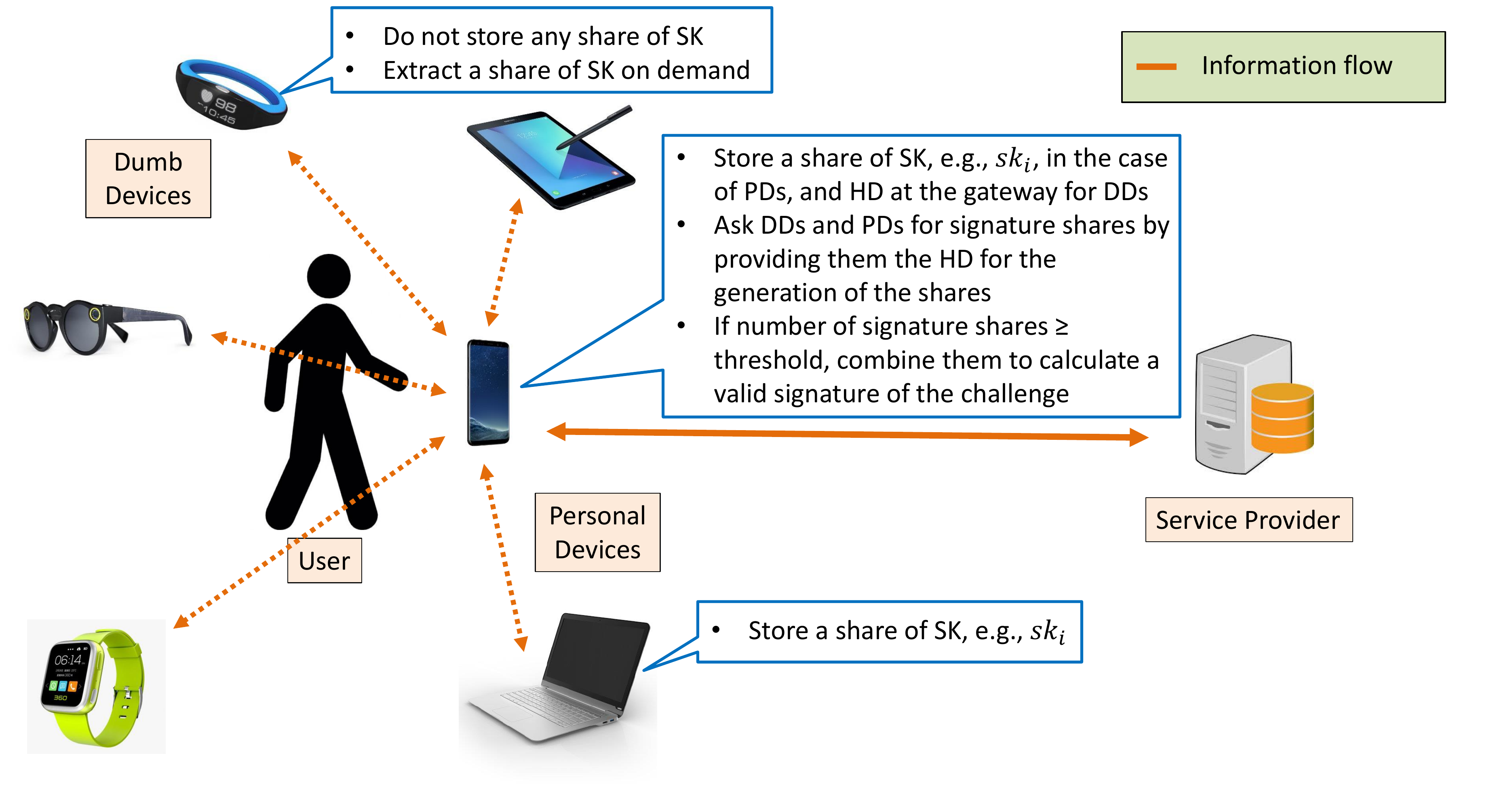}
\caption{CASE 3: FAS using threshold signature and fuzzy extractors.}
\label{Fig:case3}
\end{figure}

In the previous solution, shares of the secret key are stored in users' DDs. As these DDs usually do not have secure storage, storing sensitive data on them (i) might be undesirable and (ii) can pose a threat to security of the FAS, in general. To overcome this limitation, one option is to use Fuzzy Extractors (FEs) to allow DDs to recover their shares of the secret key, thus avoiding the storage of sensitive data on DDs (see Figure~\ref{Fig:case3}). FEs use noisy data from a source and Helper Data (HD) to recover a fixed discrete representation. Using mechanisms such as the uncoupling procedure presented in~\cite{Abidin2017}, where the binary representation bound in the fuzzy commitment is independent of the fuzzy source, it is possible to make a FE to produce a given key, producing HD which does not disclose any information about the produced key. In our case, each DD uses a FE to obtain its corresponding key share, and the HD are stored in the user's PD. During the enrolment stage, a key share and the associated HD is generated for each DD. The key share is discarded, while the HD is stored in the PD. During the authentication stage, the PD provides the DDs with their corresponding HD. Then, DDs use the collected sensory data and the provided HD to recover the corresponding key share by using the FE. This generated key share is then used to jointly sign the challenge. 

\subsubsection{Advantages}
The online generation of the key shares during the authentication stage means that \textit{key shares are not stored} at different devices, thus the security threat associated to their storage simply disappears. In addition, \textit{the stored HD is unlinked with the key shares}, thus avoiding information disclosure and improving the security of the system.

\subsubsection{Disadvantages}
This solution relies on the use of FE, where performance issues and the nature of the stored HD have to be taken into account when evaluating the risks. Although the HD is not linked to the produced key shares, \textit{the stored HD is linked to the biometrics/behaviometrics of the user}, thus providing information about the user's biometric data, which could be used to link the user amongst services, or to obtain information useful for spoofing attacks. Therefore, the HD have to be protected and stored in a secure element in the PD. There might also be some \textit{performance issues} as FEs differ from authentication methods based on fixed factors in the associated uncertainty in their outputs. They are subject to possible errors in genuine attempts (False Rejections) and impostor attempts (False Acceptances). In our case, several DDs will collaborate to generate a response, and $t+1$ of them need to successfully recover their respective share. These considerations should be kept in mind, when generating the HD, to properly decide the working point for different FEs.


\section{Conclusions and Future Work}
\label{Conclusions}
In this paper we have presented a comprehensive security and privacy analysis of a FAS, starting from a set of functional requirements. Three different approaches for a secure and privacy-friendly FAS have been analysed, integrating possession-based and behavioural authentication factors in a flexible authentication scheme based on threshold signatures. The main advantages and disadvantages of the different approaches have been analysed. Although all the three analysed solutions meet the main security and privacy requirements, we recommend the solution that combines threshold signature with fuzzy extractors, as no key material is stored at user devices. As future work, we will design a concrete protocol for a FAS that combines threshold signature with fuzzy extractors, and evaluate its performance in terms of computational complexity, communication costs, and authentication rates.





%
%
%

\bibliographystyle{IEEEtran}
\bibliography{FrictAuthBiblio}

\end{document}